 \documentclass[manuscript]{aastex}

\usepackage{bm}
\usepackage{amsmath,mathtools}
\usepackage[subnum]{cases}
\usepackage{multirow}

\shorttitle{MODIFICATION OF NLGC}
\shortauthors{QIN AND ZHANG}

\begin{document}
              \arraycolsep 0pt

\title{THE MODIFICATION OF THE NONLINEAR GUIDING CENTER THEORY}

\author{G. Qin\altaffilmark{1} and L.-H. Zhang\altaffilmark{1,2}}
\email{Gang Qin, gqin@spaceweather.ac.cn; Lihua Zhang, lhzhang@spaceweather.ac.cn}
\altaffiltext{1}{State Key Laboratory of Space Weather, Center for Space Science and
 Applied Research, Chinese Academy of Sciences, Beijing 100190, China.}
\altaffiltext{2}{College of Earth Sciences, Graduate University of Chinese Academy 
of Sciences, Beijing 100049, China.}

\begin{abstract}
We modify the NonLinear Guiding Center (NLGC) theory (Matthaeus et al. 2003) for 
perpendicular diffusion by replacing the spectral 
amplitude of the two-component model magnetic turbulence with the 2D component one
(following Shalchi 2006), and replacing the constant $a^2$, indicating the degree 
particles following magnetic field line, with a variable $a^{\prime 2}$ as a 
function of the magnetic turbulence. We combine the modified model with the 
NonLinear PArallel (NLPA) diffusion theory (Qin 2007) to solve perpendicular 
and parallel 
diffusion coefficients simultaneously. It is shown that the new model agrees better 
with simulations. Furthermore, we fit the numerical results of the new model with 
polynomials, so that parallel and perpendicular diffusion coefficients can be 
calculated directly without iteration of integrations, and many numerical 
calculations can be reduced.
\end{abstract}

\keywords{}

\section{INTRODUCTION}
Knowledge of charged energetic particles' diffusion mechanism is necessary to study 
the transport and acceleration of cosmic rays. \citet{Matthaeus2003} developed a
NonLinear Guiding Center (NLGC)
\footnote{Note that the acronyms in this paper are briefly explained in Table 
\ref{tbl:terms}.}
 theory to describe the perpendicular diffusion
coeficient, which is written as \citep{Matthaeus2003} 
\begin{equation}
\kappa_{xx}=\frac{a^2v^2}{3B_0^2}\int d^3\bm{k}\frac{S_{xx}(\bm{k})}{\frac{v}
{\lambda_{\parallel}}+k_{\bot}^2\kappa_{xx}+k_{\parallel}^2\kappa_{zz}+
\gamma(\bm{k})},\label{NLGC}
\end{equation}
where the parameter $a^2=1/3$ indicating the degree particles following Magnetic
Field Line (MFL). 
In addition, the spectral amplitudes of 
two-component model turbulence $S_{xx}(\bm{k})$ is the sum of that of 2D component 
$S_{xx}^{2D}(\bm{k})$ and slab component $S_{xx}^{slab}(\bm{k})$ \citep[e.g.,][]
{BieberEA96},
\begin{eqnarray}
&&S_{xx}(\bm{k})=S_{xx}^{2D}(\bm{k})+S_{xx}^{slab}(\bm{k})\nonumber\\
&=&S_{xx}^{\prime 2D}(k_{\bot})\frac{2k_y^2\delta(k_{\parallel})}{\pi k_{\bot}^3}
+S_{xx}^{\prime slab}(k_{\parallel})\frac{\delta(k_{\bot})}{2\pi k_{\bot}},
\end{eqnarray}
with 2D component $S_{xx}^{\prime 2D}(k_\perp)$ and slab component 
$S_{xx}^{\prime slab}(k_\parallel)$ written as
\begin{eqnarray}
&&S_{xx}^{\prime 2D}(k_{\bot})=C(\nu)\lambda_{2D}\langle b_{2D}^2\rangle\left(1+
k_{\bot}^2\lambda_{2D}^2\right)^{-\nu}  \label{eq:Sxx2D}\\
&&S_{xx}^{\prime slab}(k_{\parallel})=C(\nu)\lambda_{slab}\langle b_{slab}^2\rangle
\left(1+k_{\parallel}^2\lambda_{slab}^2\right)^{-\nu},  \label{eq:Sxxslab}
\end{eqnarray}
and $C(\nu)=\frac{1}{2\sqrt{\pi}}\frac{\Gamma(\nu)}{\Gamma(\nu-1/2)}$. 
Here, $\lambda_{slab}$ and $\lambda_{2D}$ are the spectral bend-over scales of slab
and 2D components of turbulence, respectively, and particles parallel mean free path
 is related to the parallel diffusion coefficient as
$\lambda_\parallel=3\kappa_{zz}/v$. It is considered that NLGC is the first 
perpendicular diffusion theory that agrees well with simulations and spacecraft 
observations in typical solar wind conditions \citep{ZankEA04, BieberEA04}.
Note that in Eqs. (\ref{eq:Sxx2D}) and 
(\ref{eq:Sxxslab}) the spectra of 2D and slab components, respectively, are all set
 flat in energy range with lower wavenumbers. However, \citet{MatthaeusEA2007} 
suggested that in 
energy range, $k_\perp\ll 1/\lambda_{2D}$, the spectrum of 2D component should not 
be constant.
In addition with a more general form of the 2D spectrum with energy range spectrum
 index $q$, \citet{ShalchiEA2010} found that the behavior of the spectrum in energy
range is important to determine the perpendicular diffusion. 

Furthermore, the NLGC model is an 
integral equation for perpendicular diffusion coefficient with parallel diffusion 
coefficient as an input, so it is difficult to include NLGC in a numerical model to
study energetic particles transport or acceleration.
\citet{ZankEA04} and \citet{ShalchiEA04anal} derived explicit expressions for the 
perpendicular diffusion coefficient, where parallel diffusion coefficient from
Quasilinear Theory \citep[QLT][]{Jokipii1966} can be used as input. Furthermore, 
\citet{ShalchiEA04} developed a nonlinear 
model, WNLT, to describe parallel and perpendicular diffusion simultaneously. 
But the WNLT model is complicated and difficult to use in numerical models. 

Moreover, because of nonlinear effects, QLT is not very accurate compared to the 
simulation results \cite[e.g.,][]{Qin2002}. Based on the NLGC theory, a nonlinear 
model, the NLPA theory of parallel diffusion coefficient is derived 
as \citep{Qin2007}
\begin{eqnarray}
&\kappa_{zz}&=\Bigg\{6a_x\left(\frac{\Omega}{v}\right)^2\int d^3\bm{k}
\frac{S_{xx}(\bm{k})}{B_0^2}\nonumber\\
&\times&\frac{\frac{v}{\lambda_{\parallel}}+
k_{\bot}^2\kappa_{xx}+k_{\parallel}^2\kappa_{zz}}{\Omega^2+
[\frac{v}{\lambda_{\parallel}}+k_{\bot}^2\kappa_{xx}+k_{\parallel}^2\kappa_{zz}+
\gamma(\bm{k})]^2}\Bigg\}^{-1},\label{NLPA}
\end{eqnarray}
with the parameter
\begin{equation}
a_x=\frac{1}{2}\sqrt{\frac{\tilde{E}_s}{[\xi/(1+\xi)](1/\tilde{b})+
\tilde{b}/(2\xi)}},
\end{equation}
and $\tilde{r}=2\pi r_L/\lambda_c$, $\tilde{b}=b/B_0$, $\xi=\tilde{r}/\tilde{b}$, 
$\tilde{E}_s=E_{slab}/E_{total}$, and $E_{total}=E_{slab}+E_{2D}$. Here, 
the correlation length of the slab turbulence $\lambda_c$ is related to the slab 
turbulence correlation scale $\lambda_{slab}$ as
$\lambda_c=2\pi C(\nu)\lambda_{slab}$, $r_L$ is the particle
maximum gyro-radius, $b=\left(\langle b^2_{slab}\rangle+\langle b^2_{2D}\rangle
\right)^{1/2}$, $b/B_0$ is the turbulence level, and 
$E_{slab}=\langle b^2_{slab}\rangle$ and $E_{2D}=\langle b^2_{2D}\rangle$ 
are magnetic turbulence energy in slab and 2D components, respectively.
Note that 
in equation (8) of \citet{Qin2007} there was a typo in the defination of the 
parameter $a_x$ which was corrected in \citet{Qin2013}. It is also noted that 
QLT \citep{Jokipii1966} can not be obtained from this theory in the corresponding 
limit. Again, NLPA model is an 
integral equation for parallel diffusion coefficient with perpendicular 
diffusion coefficient as an input. Furthermore, the NLGC 
and NLPA theories can be combined to get the NLGC-E model to determine the
parallel and perpendicular diffusion coefficients simultaneously \citep{Qin2007}.

In addition, \citet{Shalchi2010} developed a unified diffusion theory for 
perpendicular diffusion based on \citet[NLGC,][]{Matthaeus2003}, with the 
Fokker-Planck equation to compute the fourth-order correlations during derivation. 
Furthermore,
\citet{TautzAShalchi2011} compared the unified diffusion theory, noted as INLGC,
 (as well as NLGC) with simulations. It is shown that the unified diffusion theory,
INLGC, can be used for 3D turbulence. Moreover, INLGC automatically satisfies the 
subdiffusive result for slab turbulence ($\kappa_\perp=0$), and corresponds to NLGC 
for two-component turbulence without any additional assumptions.

In this study, for simplicity purpose, we modify the NLGC theory directly by 
replacing the spectral 
amplitude of the two-component model magnetic 
turbulence with the 2D component one and replacing the constant indicating the 
degree particles following magnetic field line with a variable of magnetic 
turbulence. We also fit the numerical results of the modified model with polynomials.
The paper is organized as follows. We discuss the modification of the NLGC theory
in section 2. The polynomial fitting of the new model is discussed in section 3.
Finally, conclusions are presented in section 4.

\section{MODIFICATION OF THE NLGC THEORY}

The NLGC theory agrees with simulation results very well in general solar
wind conditions. However, from simulation results in \citet{Qin2007}, especially in 
Figure 3 of
 \citet{Qin2007}, we find that the NLGC theory for perpendicular diffusion does not
agree simulation results well when the turbulence is nearly pure slab or pure 2D, so
it is necessary for one to modify the NLGC theory. Firstly, it is considered that
 the slab 
component of the turbulent magnetic field does not contribute directly to the 
perpendicular diffusion \citep{Shalchi2006, Shalchi2010}. Therefore, in the NLGC 
theory of the 
perpendicular diffusion coefficient equation (\ref{NLGC}), the spectral amplitude of 
the two-component model magnetic turbulence $S_{xx}(\bf{k})$ should only include the
 2D component, $S_{xx}^{2D}(\bf{k})$ \citep{Shalchi2006}. 
Secondly, we assume the degree particles 
following MFL is varied with the conditions of magnetic turbulence, so we modify the
 parameter $a^2$ in equation (\ref{NLGC}) with different forms and compare with 
simulation results in 
\citet{Qin2007}, and the best form we can get so far is
\begin{equation}
a^{\prime 2}=\left(\sqrt{\frac{\lambda_{2D}}{\lambda_{slab}}}\frac{E_{total}}
{E_{slab}}+\frac{4}{3}\frac{E_{total}}{E_{2D}}\right)^{-1}.
\label{eq:amod}
\end{equation}
Therefore, by replacing $S_{xx}(\bf{k})$ and $a^2$ with $S_{xx}^{2D}(\bf{k})$ and 
$a^{\prime 2}$, respectively, in equation (\ref{NLGC}), we get a modified NLGC 
theory for perpendicular diffusion coefficient,
\begin{eqnarray}
\kappa_{xx}&=&\frac{a^{\prime 2}v^2}{3}\int dk_{\bot}2C(\nu)\lambda_{2D}
\frac{\langle b_{2D}^2\rangle}{B_0^2}\left(1+k_{\bot}^2\lambda_{2D}^2\right)^{-\nu}
\nonumber\\
&\times&\frac{1}{\frac{v}{\lambda_{\parallel}}+k_{\bot}^2\kappa_{xx}+
\gamma(\bm{k})}.\label{NLGC-N}
\end{eqnarray}
Here, symmetric 2D component turbulence is assumed.
In addition, the 2D spectrum of Eq. (\ref{eq:Sxx2D}) with energy range spectrum 
index $q=0$ is used in 
order to compare with simulations in \cite{Qin2007}. In the future, a more general
form of 2D spectrum with $q\ne 0$ can be used.
In the following, we note this modified model as NLGC-N. In addition, we combine the
NLGC-N and NLPA models, the equations (\ref{NLGC-N}) and (\ref{NLPA}), respectively,
 to get an NLGCE-N model.
Next we compare the numerical results of NLGCE-N model with that of NLGC-E and the 
simulation results from \citet{Qin2007}. Here, $\gamma(\bm{k})$ is chosen to be $0$ 
for static magnetic turbulence, and $\nu$ is chosen to be $5/6$.
In addition, we can also define a parameter, the perpendicular mean free path 
$\lambda_\perp\equiv 3\kappa_{zz}/v$ for simplicity purpose.

\begin{figure}
  \centering
  \includegraphics[width=.7\textwidth]{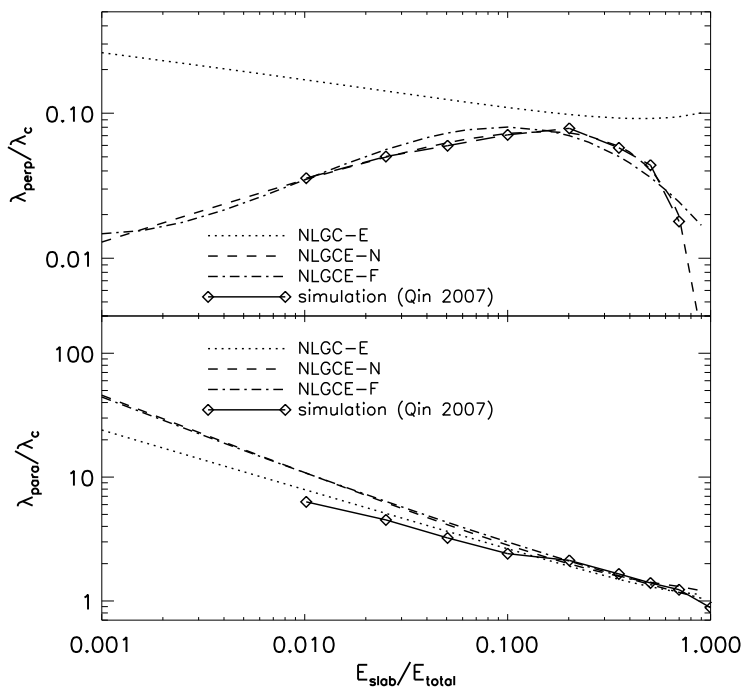}
  \caption{Top and bottom panels show perpendicular and parallel mean free paths,
respectively, as a function of $E_{slab}/E_{total}$, with $r_L/\lambda_c=0.048$,
$b/B_0=1$ and $\lambda_{2D}/\lambda_{slab}=0.1$. Diamonds are from the simulations 
in \citet{Qin2007}, Dotted, dashed, and dashed-dotted lines indicate results from 
NLGC-E, NLGCE-N, and NLGCE-F, respectively.}
  \label{fig:eslab}
\end{figure}

Top and bottom panels of Figure \ref{fig:eslab} show perpendicular and parallel mean
 free paths,
respectively, as a function of $E_{slab}/E_{total}$, with $r_L/\lambda_c=0.048$,
$b/B_0=1$ and $\lambda_{2D}/\lambda_{slab}=0.1$. Diamonds are from the simulations
in \citet{Qin2007}, Dotted, dashed, and dashed-dotted lines indicate results from
NLGC-E, NLGCE-N, and NLGCE-F, respectively. Later we will study NLGCE-F, which is 
the polynomial fitting of NLGCE-N. The simulation results in Figure \ref{fig:eslab}
are obtained from the Figure 3 of \citet{Qin2007}. As already shown in 
\citet{Qin2007}, the perpendicular diffusion coefficient from NLGC-E agrees well 
with simulation results when $E_{slab}/E_{total}\sim 0.2$, but it does not agree
well with simulations when $E_{slab}/E_{total}\ll 0.2$ or 
$E_{slab}/E_{total}\to 1$. However, the perpendicular diffusion coefficient from the
new model, NLGCE-N, agrees very well for the whole range of $E_{slab}/E_{total}$,
from $E_{slab}/E_{total}\ll 0.2$ to $E_{slab}/E_{total}\to 1$. At the same time,
the parallel diffusion coefficient from NLGCE-N generally agrees well with 
simulations. Although with $E_{slab}/E_{total}\ll 0.02$, the NLGCE-N is relatively 
worse than NLGC-E, the results of NLGCE-N are still acceptable.

\begin{figure}
  \centering
  \includegraphics[width=.7\textwidth]{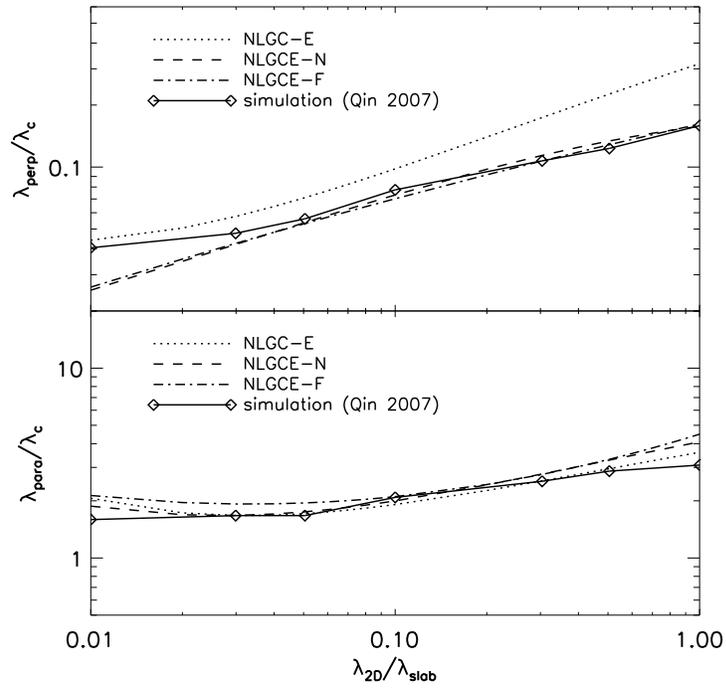}
  \caption{Similar as Figure \ref{fig:eslab}, except that x-axis is 
$\lambda_{2D}/\lambda_{slab}$ with $E_{slab}/E_{total}=0.2$.}
  \label{fig:lambda2d}
\end{figure}

Figure \ref{fig:lambda2d} is similar to Figure \ref{fig:eslab}, except that x-axis 
is $\lambda_{2D}/\lambda_{slab}$ with $E_{slab}/E_{total}=0.2$, and that the 
simulation results are obtained from the Figure 4 of \citet{Qin2007}. From the 
bottom panel of Figure \ref{fig:lambda2d} we can see that both NLGC-E and NLGCE-N
agree well with simulations in parallel diffusion. However, from the top panel of
Figure \ref{fig:lambda2d} the two models are different in perpendicular diffusion.
NlGC-E agrees well with simulations in perpendicular 
diffusion coefficient with $\lambda_{2D}/\lambda_{slab}\sim 0.01$, but the agreement
becomes worse as $\lambda_{2D}/\lambda_{slab}$ increases. On the other hand, to 
compare to simulations, NLGCE-N is worse than NLGC-E in perpendicular diffusion with 
$\lambda_{2D}/\lambda_{slab}\lesssim 0.02$, but it is better than NLGC-E with
$\lambda_{2D}/\lambda_{slab}\gtrsim 0.03$. Generally, NLGCE-N agrees better with
simulations than NLGC-E in perpendicular diffusion.

\begin{figure}
  \centering
  \includegraphics[width=.7\textwidth]{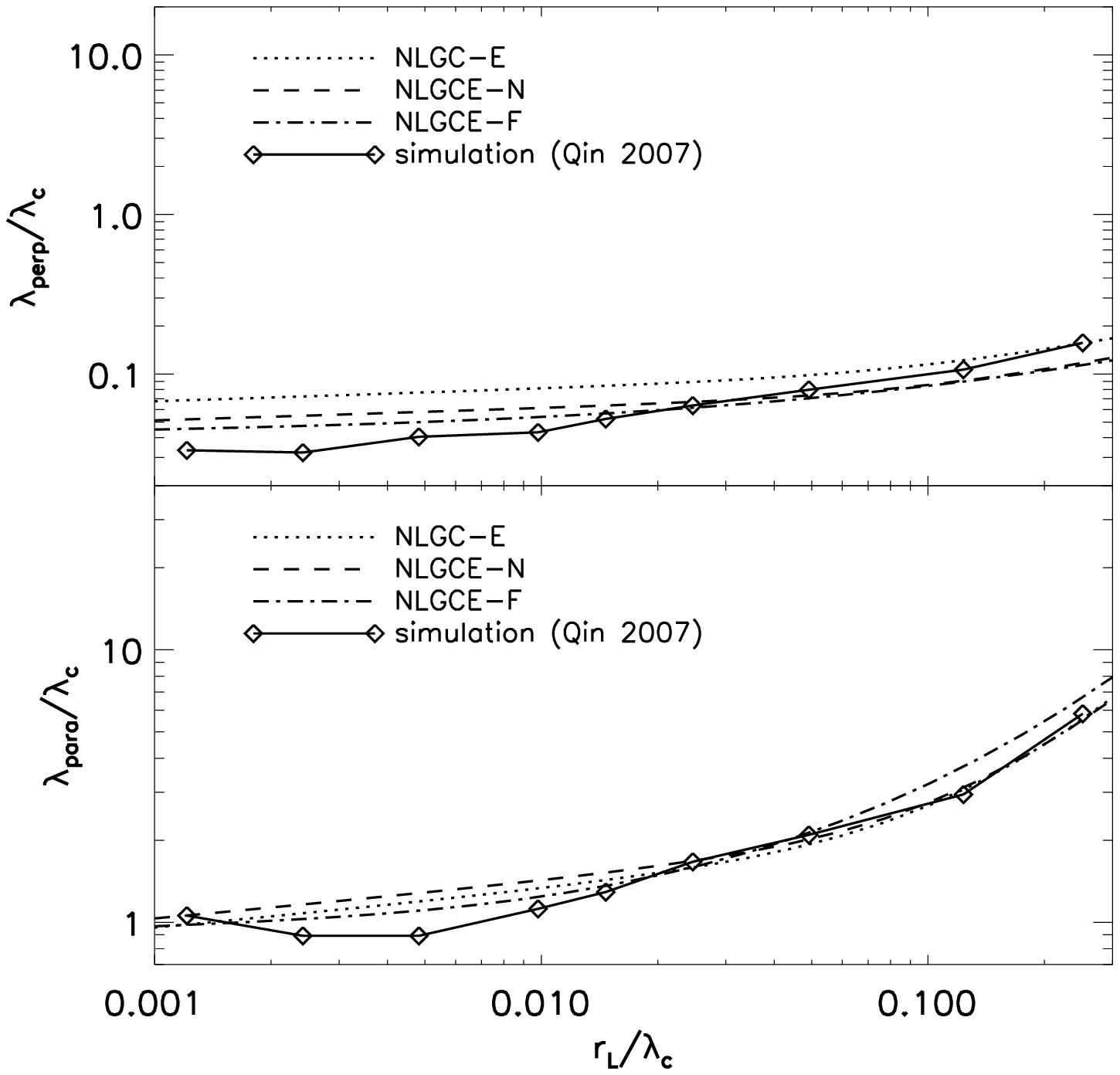}
  \caption{Similar as Figure \ref{fig:eslab}, except that x-axis is 
$r_L/\lambda_c$ with $E_{slab}/E_{total}=0.2$.}
  \label{fig:rl}
\end{figure}

\begin{figure}
  \centering
  \includegraphics[width=.7\textwidth]{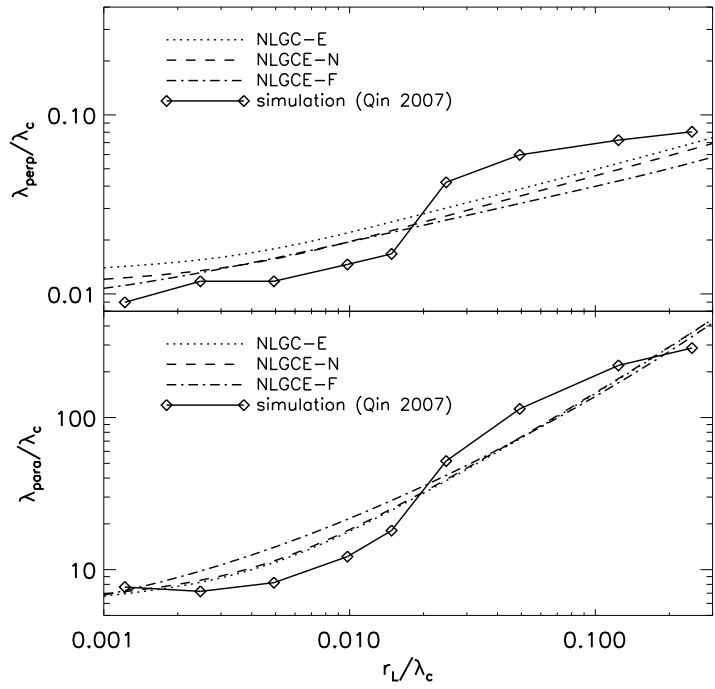}
  \caption{Similar as Figure \ref{fig:eslab}, except that x-axis is
$r_L/\lambda_c$ with $E_{slab}/E_{total}=0.2$ and $b/B_0=0.2$.}
  \label{fig:rlb0_2}
\end{figure}

Figure \ref{fig:rl} is similar to Figure \ref{fig:eslab}, except that x-axis is
$r_L/\lambda_c$ with $E_{slab}/E_{total}=0.2$, and that the simulation results are 
obtained from the Figure 1 of \citet{Qin2007}. From the Figure \ref{fig:rl} we can 
see that both NLGC-E and NLGCE-N agree well with simulations.
In addition, Figure \ref{fig:rlb0_2} is similar to Figure \ref{fig:eslab}, except 
that x-axis is $r_L/\lambda_c$ with $E_{slab}/E_{total}=0.2$ and $b/B_0=0.2$, and
that the simulation results are obtained from the Figure 2 of \citet{Qin2007}.
Again, from the Figure \ref{fig:rlb0_2} we can see that both NLGC-E and NLGCE-N 
agree well with simulations.

Therefore, we show that the new NLGCE-N model is improved compared to the NLGC-E
model, especially when the magnetic turbulence is nearly pure slab or 2D.

\begin{figure}
  \centering
  \includegraphics[width=.7\textwidth]{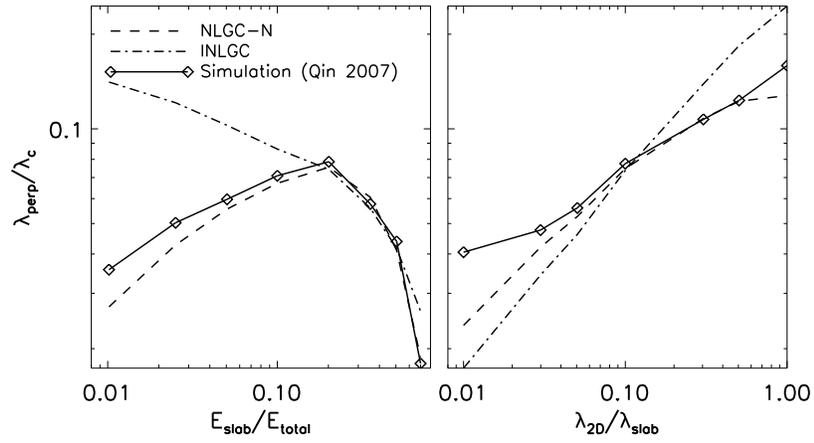}
  \caption{Left and right panels are similar as top panels of Figure \ref{fig:eslab}
and Figure \ref{fig:lambda2d}, respectively, except that dashed and dashed-dotted
lines  indicate results from NLGC-N and INLGC, respectively.}
  \label{fig:scompare}
\end{figure}

 In addition, we compare the modified model NLGC-N with the INLGC model 
\citep{Shalchi2010,TautzAShalchi2011}. Since the NLGC-N and INLGC are models for
perpendicular diffusion with parallel diffusion coefficients as input, we use 
$\kappa_\parallel$ from the simulation results in \citet{Qin2007} as input. Left and
right panels of Figure \ref{fig:scompare} are similar as top panels of Figure 
\ref{fig:eslab} and Figure \ref{fig:lambda2d}, respectively, except that dashed and 
dashed-dotted lines indicate results from NLGC-N and INLGC, respectively.
From left panel of the figure we can see that when $0.1<E_{slab}/E_{total}\lesssim 
0.5$, both NLGC-N and INLGC agree well with simulations. But when 
$E_{slab}/E_{total}\lesssim 0.1$ or $E_{slab}/E_{total}>0.5$, NLGC-N agrees better 
with simulations than INLGC. Furthermore, from right 
panel of the figure we can see that when $\lambda_{2D}/\lambda_{slab}\sim 0.1$, both
NLGC-N and INLGC agree well with simulations, but when 
$\lambda_{2D}/\lambda_{slab}\not\approx 0.1$, NLGC-N agrees better with simulations 
than INLGC. 

\section{POLYNOMIAL FITTING OF NLGCE-N}
Although the new NLGCE-N model agrees with simulations well, but numerous iteration 
of integrations for both of the equations are needed to solve it numerically. In 
order to simplify the numerical calculations needed using the model to study 
energetic particles transport or acceleration, we fit the equations of NLGCE-N model
 with polynomials in parameters of magnetic field and particles, i.e., 
$r_L/\lambda_{slab}$, $E_{slab}/E_{total}$, $b^2/B^2_0$, and 
$\lambda_{slab}/\lambda_{2D}$, or
\begin{equation}
\ln{\frac{\lambda_{\alpha}}{\lambda_{slab}}}=\sum_{i=0}^{n_{\alpha 1}}a_i^\alpha\left(\ln{
\frac{r_L}{\lambda_{slab}}}\right)^i,
\end{equation}
with
\begin{eqnarray*}
&&a_i^\alpha=\sum_{j=0}^{n_{\alpha 2}}b_{i,j}^\alpha\left(\ln{\frac{E_{slab}}{E_{total}}}
\right)^j\\
&&b_{i,j}^\alpha=\sum_{k=0}^{n_{\alpha 3}}c_{i,j,k}^\alpha\left(\ln{\frac{b^2}{B_0^2}}
\right)^k\\
&&c_{i,j,k}^\alpha=\sum_{l=0}^{n_{\alpha 4}}d_{i,j,k,l}^\alpha\left(\ln{\frac{\lambda_{slab}}
{\lambda_{2D}}}\right)^l
\end{eqnarray*}
where $\alpha$ indicates $\perp$ or $\parallel$. Note that this formula is not
valid in pure 2D turbulence. By fitting the
numerical results of NLGCE-N with the polynomials in wide ranges of parameters,
$r_L/\lambda_{slab}$, $E_{slab}/E_{total}$, $b^2/B^2_0$, and 
$\lambda_{slab}/\lambda_{2D}$
as shown in Table \ref{tbl-1}, we get $\Pi_{i=1}^4(n_{\alpha i}+1)$ coefficients
$d_{i,j,k,l}^\alpha$ for either parallel or perpendicular diffusion. It is noted 
that with larger values of $n_{\alpha i}$, we can get a fitting formula with higher
accuracy, but with more fitting parameters $d_{i,j,k,l}^\alpha$ needed. So we have 
to balance between accuracy and the simplicity of the fitting formula. We tried 
fitting formulae with different set of $n_{\alpha i}$ and compared the results with 
the NLGCE-N carefully (not shown), and we found that with $n_{\alpha 1}=5$, 
$n_{\alpha 2}=3$, $n_{\alpha 3}=3$, and $n_{\alpha 4}=2$ for both parallel and 
perpendicular diffusion, we can get a fitting formula with good accuracy and 
acceptable calculation scale. This way, we can directly calculate the
parallel 
and perpendicular diffusion coefficients without iteration of integrations, and the
polynomial fitting results of parallel and perpendicular diffusion coefficients are
called NLGCE-F. 
The $\Pi_{i=1}^4(n_{\alpha i}+1)=6\times 4\times 4\times 3=288$ coefficients 
$d_{i,j,k,l}^\alpha$ for parallel ($\alpha=\parallel$) and perpendicular 
($\alpha=\perp$) diffusion in the NLGCE-F model are shown in Tables 
\ref{tbl:dijklpar} and \ref{tbl:dijklperp}, respectively. 

\begin{figure}
  \centering
  \includegraphics[width=.7\textwidth]{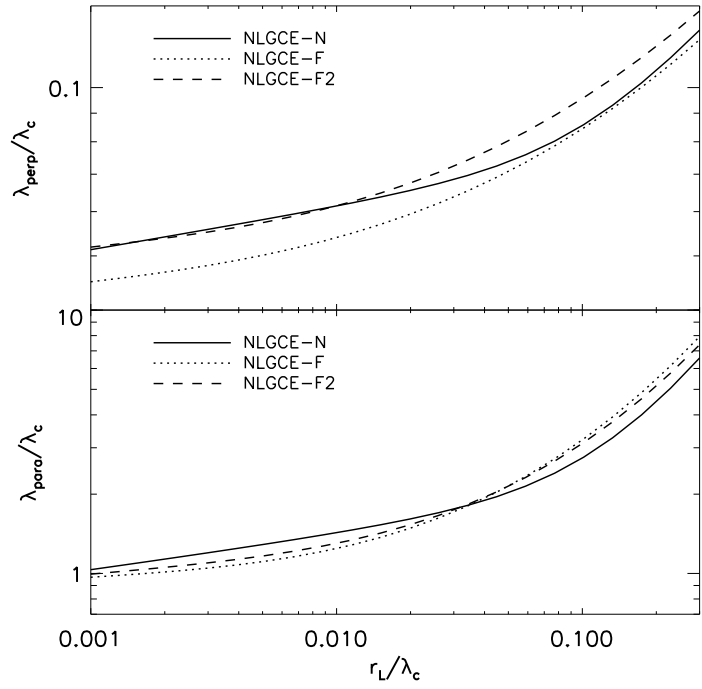}
  \caption{Similar as Figure \ref{fig:rl}, except that solid, dotted, and dashed
lines indicate results from NLGCE-N, NLGCE-F, and NLGCE-F2, respectively.}
  \label{fig:fitcompare}
\end{figure}

As an example to show the NLGCE-F's acceptable accuracy with controllable
calculation scale, in Figure
\ref{fig:fitcompare} we show the comparison between the NLGCE-F and the results of 
a new fitting formula {\bf with $n_{\alpha 1}=6$, $n_{\alpha 2}=4$, 
$n_{\alpha 3}=4$, and $n_{\alpha 4}=3$,} which is noted as NLGCE-F2. 
Figure \ref{fig:fitcompare} is Similar as Figure \ref{fig:rl}, except that solid, 
dotted, and dashed lines indicate results
from NLGCE-N, NLGCE-F, and NLGCE-F2, respectively.
From top panel of the figure we can see that for perpendicular diffusion, 
when $r_L/\lambda_c< 0.03$, 
NLGCE-F2 agrees better with NLGCE-N than NLGCE-F, but when $r_L/\lambda_c> 0.03$,
NLGCE-F agrees better with NLGCE-N than NLGCE-F2. However, from bottom panel of the 
figure we can see that for parallel diffusion, NLGCE-F
and NLGCE-F2 agree very well with each other in the range $0.001\lesssim 
r_L/\lambda_c\lesssim 0.3$. With comparisons including other variable ranges (not 
shown) we
found that, generally speaking, relative to NLGCE-F2, NLGCE-F is acceptable in
agreement with NLGCE-N. But with NLGCE-F2, the number of coefficients 
$d_{i,j,k,l}^\alpha$ for ether parallel or perpendicular diffusion 
{\bf is $7\times5\times5\times4=700$.}
Furthermore, in order to show the 
agreement between the model NLGCE-N and its polynomial fitting NLGCE-F, in Figures 
\ref{fig:eslab}-\ref{fig:rlb0_2}, we 
plot the results of NLGCE-F with dashed-dotted lines. From the figures we can see 
that NLGCE-F agrees with NLGCE-N relatively well.

\section{CONCLUSIONS}
In this paper, we modified the NLGC model, equation (\ref{NLGC}), which determines 
particles perpendicular diffusion, by replacing the spectral amplitude of the 
two-component model magnetic turbulence $S_{xx}(\bm{k})$ with the 2D component 
one, $S_{xx}^{2D}(\bf{k})$ \citep{Shalchi2006}, and replacing the parameter $a^2$ 
with $a^{\prime 2}$ which is a function of $E_{slab}/E_{total}$ and $\lambda_{2D}/
\lambda_{slab}$, to get a new model NLGC-N for perpendicular diffusion. To combine 
NLGC-N with NLPA, the model for parallel diffusion, we get
a model NLGCE-N, which can be solved simultaneously to describe perpendicular and 
parallel diffusion. In addition, we show that NLGCE-N agrees better with simulations
 than NLGC-E, which is the combination of NLGC and NLPA. Furthermore, we fit the 
numerical results of NLGCE-N with the polynomials in wide ranges of parameters
$r_L/\lambda_{slab}$, $E_{slab}/E_{total}$, $b^2/B_0^2$, and 
$\lambda_{slab}/\lambda_{2D}$, to get a new model NLGCE-F. 
So that we can directly calculate parallel and 
perpendicular diffusion coefficients simultaneously without iteration of 
integrations. Therefore, much numerical calculations would be saved to study 
diffusion coefficients.

It is also noted that when $E_{slab}/E_{total}\lesssim 0.1$ or 
$\lambda_{2D}/\lambda_{slab}\not\approx 0.1$, the modified model NLGC-N agrees 
better with simulations than the unified diffusion theory (INLGC). In addition, the 
modified model NLGC-N is very similar to the two component limit of the unified
diffusion theory, with a major difference, i.e., in INLGC a constant 
parameter $a^2=1/3$ is used. So it is 
suggested that the unified diffusion theory can also adopt the similar modification 
of parameter $a^2$, Eq. (\ref{eq:amod}), as in this paper.

In the future, we would compare our models with simulations with general forms 
of 
2D component with energy range spectrum index $q\ne 0$. In addition, we would use 
the model NLGCE-F to study transport of energetic particles in solar wind, including
 solar energetic particles, anomalous cosmic rays, or galactic cosmic rays 
\citep[e.g.,][]{QinEA06, ZhaoEA14}. Furthermore, we put the FORTRAN code for NLGCE-F
 with the data of the coefficients $d_{i,j,k,l}^\alpha$ online at 
http://www.qingang.org.cn/code/NLGCE-F to be freely downloaded and used by anybody.

\acknowledgments

We are partly supported by grants NNSFC 41374177, NNSFC 41125016, NNSFC 41304135,
and the Specialized Research Fund for State Key Laboratories of China.
The computations were performed by Numerical Forecast Modeling R\&D and VR System of
  State Key Laboratory of Space Weather and Special HPC work stand of Chinese
Meridian Project.
We appreciate the reviews provided by the referee.

\clearpage
\begin{table}[ht]
\begin{center}
\caption{The notation of terms.\label{tbl:terms}}
\begin{tabular}{|c|c|c|c|c|}
\tableline
Terms & Explanation &Output&Input & Papers\\
\tableline
NLGC & NonLinear Guiding Center theory &$\kappa_\perp$&
$\kappa_\parallel$&M03\tablenotemark{a}\\
\tableline
WNLT & Weakly NonLinear Theory &$\kappa_\perp$, $\kappa_\parallel$&&
S04B\tablenotemark{b}\\
\tableline
NLPA & NonLinear PArallel diffusion theory &$\kappa_\parallel$&$\kappa_\perp$& 
Q07\tablenotemark{c}\\
\tableline
NLGC-E & Combination of NLGC and NLPA& $\kappa_\perp$, $\kappa_\parallel$&& 
Q07\tablenotemark{c}\\
\tableline
INLGC & Unified diffusion theory &$\kappa_\perp$&$\kappa_\parallel$& 
S10\tablenotemark{d}, TA11\tablenotemark{e}\\
\tableline
NLGC-N & Modification of NLGC &$\kappa_\perp$&$\kappa_\parallel$& 
TP\tablenotemark{f}\\
\tableline
NLGCE-N &  Combination of NLGC-N and NLPA &$\kappa_\perp$, $\kappa_\parallel$&& 
TP\tablenotemark{f}\\
\tableline
NLGCE-F & Polynomial fitting of NLGCE-N &$\kappa_\perp$, $\kappa_\parallel$&& 
TP\tablenotemark{f}\\
\tableline
\end{tabular}
\tablenotetext{a}{\citet{Matthaeus2003}}
\tablenotetext{b}{\citet{ShalchiEA04}}
\tablenotetext{c}{\citet{Qin2007}}
\tablenotetext{d}{\citet{Shalchi2010}}
\tablenotetext{e}{Term noted in \citet{TautzAShalchi2011}}
\tablenotetext{f}{This paper}
\end{center}
\end{table}

\clearpage
\begin{table}[ht]
\begin{center}
\caption{The range of the four variences.\label{tbl-1}}
\begin{tabular}{|c|c|c|c|}
\tableline
$\lambda_{slab}/\lambda_{2D}$ & $E_{slab}/E_{total}$ & $b^2/B_0^2$ & 
$r_L/\lambda_{slab}$\\
\tableline
$1\sim10^3$ & $10^{-3}\sim 0.85$ & $10^{-4}\sim 10^2$ & $10^{-5}\sim6.3$\\
\tableline
\end{tabular}
\end{center}
\end{table}

\clearpage
 \begin{table}
 \caption{The coefficients $d_{i,j,k,l}^\alpha$ for parallel diffusion in the
NLGCE-F model.
\label{tbl:dijklpar}}
 \tiny{
 \begin{tabular}{|c|c|c|c|c|c|c|}
 \hline
(j k l) & i=0 & i=1 & i=2 & i=3 & i=4 & i=5 \\
 \hline
(0 0 0) &   0.23553875E+01 &   0.13029025E+01 &   0.13427579E+00 &  -0.20846185E-01 &  -0.43999458E-02 &  -0.19732612E-03 \\
(0 0 1) &  -0.66243965E-02 &   0.77768482E-01 &   0.41205730E-01 &   0.90071283E-02 &   0.90300755E-03 &   0.33110445E-04 \\
(0 0 2) &   0.21922888E-02 &  -0.39684652E-02 &  -0.17124410E-02 &  -0.94109440E-04 &  -0.11972348E-06 &  -0.45759483E-07 \\
(0 1 0) &  -0.14247343E+01 &  -0.17294922E-01 &   0.50330910E-01 &   0.22555091E-02 &  -0.54744446E-03 &  -0.36329187E-04 \\
(0 1 1) &   0.29375712E-01 &  -0.10262387E-01 &  -0.11404086E-01 &  -0.16183987E-02 &  -0.55537309E-04 &   0.92677294E-06 \\
(0 1 2) &  -0.10823704E-01 &  -0.31550348E-02 &   0.10669635E-02 &   0.34794916E-03 &   0.30278040E-04 &   0.82627464E-06 \\
(0 2 0) &   0.96583711E-01 &  -0.12322827E-01 &  -0.14806351E-01 &  -0.19674998E-02 &  -0.47832394E-04 &   0.22621783E-05 \\
(0 2 1) &  -0.16641267E-01 &  -0.38108147E-02 &   0.15277614E-02 &   0.53316833E-03 &   0.49867082E-04 &   0.14719990E-05 \\
(0 2 2) &   0.24760055E-02 &   0.75282162E-03 &  -0.23891975E-03 &  -0.11721492E-03 &  -0.13940055E-04 &  -0.51898963E-06 \\
(0 3 0) &   0.11152796E-01 &   0.31889965E-03 &  -0.23482006E-02 &  -0.47383322E-03 &  -0.30607870E-04 &  -0.57224705E-06 \\
(0 3 1) &  -0.17297051E-02 &  -0.54967106E-03 &   0.19481087E-03 &   0.72844701E-04 &   0.72128259E-05 &   0.22722512E-06 \\
(0 3 2) &   0.28333003E-03 &   0.10946972E-03 &  -0.23059951E-04 &  -0.12751296E-04 &  -0.15280551E-05 &  -0.56788270E-07 \\
(1 0 0) &  -0.14786874E+00 &   0.35413822E+00 &  -0.23571521E-01 &  -0.29883137E-01 &  -0.41108264E-02 &  -0.16630590E-03 \\
(1 0 1) &  -0.33315220E+00 &  -0.19457101E+00 &   0.48492938E-01 &   0.19439119E-01 &   0.20217803E-02 &   0.68237390E-04 \\
(1 0 2) &  -0.70191117E-02 &   0.16863137E-01 &  -0.42473573E-02 &  -0.20407538E-02 &  -0.22445321E-03 &  -0.77342338E-05 \\
(1 1 0) &  -0.12645333E+00 &  -0.57381385E-01 &   0.19660793E-01 &   0.81465626E-02 &   0.93001667E-03 &   0.34019143E-04 \\
(1 1 1) &  -0.74499928E-01 &   0.21886029E-01 &   0.78026833E-02 &  -0.90778341E-03 &  -0.27852845E-03 &  -0.13910525E-04 \\
(1 1 2) &   0.88176328E-02 &  -0.24607326E-02 &  -0.16399476E-02 &  -0.65717348E-04 &   0.18575402E-04 &   0.12449253E-05 \\
(1 2 0) &   0.32714354E-01 &  -0.96871348E-02 &  -0.88372387E-02 &  -0.10864091E-02 &   0.19497311E-06 &   0.31551668E-05 \\
(1 2 1) &  -0.87371981E-02 &   0.23907528E-02 &  -0.84712003E-03 &  -0.58296444E-03 &  -0.80491321E-04 &  -0.32954014E-05 \\
(1 2 2) &   0.20518127E-02 &   0.25128221E-03 &  -0.21217125E-04 &   0.70656292E-07 &   0.91585353E-06 &   0.56694614E-07 \\
(1 3 0) &   0.38644947E-02 &   0.50051530E-03 &  -0.11014859E-02 &  -0.31141696E-03 &  -0.28640974E-04 &  -0.86437844E-06 \\
(1 3 1) &   0.99272926E-04 &   0.56214724E-04 &  -0.18481718E-03 &  -0.55454476E-04 &  -0.54975212E-05 &  -0.18100547E-06 \\
(1 3 2) &   0.87113797E-04 &   0.41654101E-04 &   0.12554768E-04 &   0.57474926E-06 &  -0.11142497E-06 &  -0.82310051E-08 \\
(2 0 0) &   0.12726928E+00 &   0.60497042E-01 &  -0.11952481E-01 &  -0.71946142E-02 &  -0.90296336E-03 &  -0.35119674E-04 \\
(2 0 1) &  -0.12984315E+00 &  -0.42695185E-01 &   0.14716482E-01 &   0.47345284E-02 &   0.44192205E-03 &   0.13655324E-04 \\
(2 0 2) &   0.76251665E-02 &   0.42709723E-02 &  -0.15275348E-02 &  -0.53455363E-03 &  -0.50940514E-04 &  -0.15623067E-05 \\
(2 1 0) &  -0.31863123E-01 &  -0.19815411E-01 &   0.52020996E-02 &   0.28355253E-02 &   0.35739999E-03 &   0.13896490E-04 \\
(2 1 1) &  -0.16244229E-01 &   0.93012640E-02 &   0.16637012E-02 &  -0.68382140E-03 &  -0.13628576E-03 &  -0.62450050E-05 \\
(2 1 2) &   0.27835739E-02 &  -0.97187943E-03 &  -0.47272017E-03 &   0.79149520E-05 &   0.95241782E-05 &   0.53642837E-06 \\
(2 2 0) &   0.46502480E-02 &  -0.32396413E-02 &  -0.15041494E-02 &  -0.75760531E-04 &   0.16825701E-04 &   0.12521577E-05 \\
(2 2 1) &   0.21825930E-03 &   0.12337052E-02 &  -0.50023564E-03 &  -0.24279710E-03 &  -0.29501141E-04 &  -0.11178305E-05 \\
(2 2 2) &   0.33644525E-03 &   0.26880995E-04 &   0.20519435E-04 &   0.76226242E-05 &   0.90872206E-06 &   0.33814742E-07 \\
(2 3 0) &   0.58192239E-03 &   0.61367478E-04 &  -0.17658762E-03 &  -0.52589587E-04 &  -0.51889990E-05 &  -0.16995503E-06 \\
(2 3 1) &   0.26078224E-03 &   0.54482841E-04 &  -0.84274690E-04 &  -0.24125560E-04 &  -0.22392916E-05 &  -0.68572929E-07 \\
(2 3 2) &  -0.31587447E-05 &   0.88101241E-05 &   0.77940884E-05 &   0.12632533E-05 &   0.62626899E-07 &   0.32419583E-09 \\
(3 0 0) &   0.94294495E-02 &   0.37121738E-02 &  -0.96294274E-03 &  -0.49755041E-03 &  -0.60494434E-04 &  -0.23180297E-05 \\
(3 0 1) &  -0.10376127E-01 &  -0.26461773E-02 &   0.11205302E-02 &   0.31769572E-03 &   0.27403681E-04 &   0.78509192E-06 \\
(3 0 2) &   0.86310176E-03 &   0.27169634E-03 &  -0.12254550E-03 &  -0.35939188E-04 &  -0.30618079E-05 &  -0.83621815E-07 \\
(3 1 0) &  -0.25905794E-02 &  -0.16267199E-02 &   0.44903253E-03 &   0.24949716E-03 &   0.31811253E-04 &   0.12472213E-05 \\
(3 1 1) &  -0.93864940E-03 &   0.78226701E-03 &   0.65698746E-04 &  -0.76664572E-04 &  -0.13261908E-04 &  -0.58433454E-06 \\
(3 1 2) &   0.20464969E-03 &  -0.79798598E-04 &  -0.31340467E-04 &   0.26263592E-05 &   0.96852467E-06 &   0.50155474E-07 \\
(3 2 0) &   0.17937967E-03 &  -0.26284951E-03 &  -0.77522141E-04 &   0.32019542E-05 &   0.19907745E-05 &   0.11330622E-06 \\
(3 2 1) &   0.14379785E-03 &   0.11350385E-03 &  -0.52325278E-04 &  -0.22296259E-04 &  -0.25628334E-05 &  -0.93609758E-07 \\
(3 2 2) &   0.14439318E-04 &   0.13040934E-05 &   0.28224994E-05 &   0.82772388E-06 &   0.83029680E-07 &   0.27175665E-08 \\
(3 3 0) &   0.26685171E-04 &   0.27702134E-05 &  -0.91601637E-05 &  -0.29156552E-05 &  -0.30693882E-06 &  -0.10715202E-07 \\
(3 3 1) &   0.30442128E-04 &   0.55124161E-05 &  -0.77587641E-05 &  -0.21593223E-05 &  -0.19452234E-06 &  -0.57567222E-08 \\
(3 3 2) &  -0.13835817E-05 &   0.65511747E-06 &   0.75405027E-06 &   0.12656149E-06 &   0.65449343E-08 &   0.47775400E-10 \\
 \hline
 \end{tabular}
 }
 \end{table}
\clearpage

 \begin{table}
 \caption{The coefficients $d_{i,j,k,l}^\alpha$ for perpendicular diffusion in the
NLGCE-F model.
\label{tbl:dijklperp}}
 \tiny{
 \begin{tabular}{|c|c|c|c|c|c|c|}
 \hline
(j k l) & i=0 & i=1 & i=2 & i=3 & i=4 & i=5 \\
 \hline
(0 0 0) &  -0.20782397E+01 &   0.80542173E+00 &   0.43509413E-01 &  -0.21765985E-01 &  -0.34880893E-02 &  -0.14567092E-03 \\
(0 0 1) &  -0.46386191E+00 &  -0.98573919E-01 &   0.55008025E-02 &   0.60611018E-02 &   0.81158994E-03 &   0.32584425E-04 \\
(0 0 2) &  -0.20027545E-01 &   0.77767273E-02 &  -0.17837106E-03 &  -0.34531413E-03 &  -0.48546308E-04 &  -0.20284890E-05 \\
(0 1 0) &   0.32932663E-01 &   0.55078257E-01 &   0.57859672E-02 &  -0.98263229E-02 &  -0.15658018E-02 &  -0.64502352E-04 \\
(0 1 1) &   0.79801317E-01 &  -0.16056809E-01 &  -0.43059925E-02 &   0.15297052E-02 &   0.29837887E-03 &   0.13184386E-04 \\
(0 1 2) &  -0.96094000E-02 &  -0.70750559E-03 &   0.71313130E-04 &  -0.14638662E-03 &  -0.27273546E-04 &  -0.12478895E-05 \\
(0 2 0) &   0.45254433E-01 &  -0.15503612E-03 &  -0.58120004E-02 &  -0.95055489E-03 &  -0.41356187E-04 &   0.75481060E-07 \\
(0 2 1) &  -0.12156453E-01 &  -0.14601653E-02 &   0.16977702E-02 &   0.48775342E-03 &   0.45538505E-04 &   0.14225158E-05 \\
(0 2 2) &   0.14840409E-02 &   0.42407064E-03 &  -0.45749343E-04 &  -0.31128841E-04 &  -0.36845885E-05 &  -0.13458210E-06 \\
(0 3 0) &   0.32974684E-02 &  -0.60751030E-03 &  -0.81092420E-03 &  -0.65981454E-04 &   0.46104859E-05 &   0.42766278E-06 \\
(0 3 1) &  -0.10773557E-02 &  -0.29570722E-03 &   0.43219766E-04 &   0.13131432E-05 &  -0.13053938E-05 &  -0.84558084E-07 \\
(0 3 2) &   0.12302544E-03 &   0.56007217E-04 &   0.10139352E-04 &   0.24540202E-05 &   0.33361447E-06 &   0.14524518E-07 \\
(1 0 0) &  -0.11101067E+01 &   0.31228772E+00 &  -0.13177282E-01 &  -0.22255326E-01 &  -0.30794494E-02 &  -0.12445252E-03 \\
(1 0 1) &  -0.27352296E+00 &  -0.13632322E+00 &   0.95260785E-02 &   0.80307990E-02 &   0.96089118E-03 &   0.35198094E-04 \\
(1 0 2) &   0.12818726E-01 &   0.96450258E-02 &  -0.23937203E-02 &  -0.11687264E-02 &  -0.13273061E-03 &  -0.47474106E-05 \\
(1 1 0) &  -0.10207300E+00 &   0.11063794E-01 &   0.55260570E-02 &  -0.14631952E-02 &  -0.28667936E-03 &  -0.12195614E-04 \\
(1 1 1) &  -0.89940234E-02 &   0.15794357E-01 &   0.10450265E-01 &   0.23173513E-02 &   0.20829148E-03 &   0.66426911E-05 \\
(1 1 2) &   0.32828567E-02 &  -0.10346441E-02 &  -0.10776541E-02 &  -0.22747122E-03 &  -0.19402946E-04 &  -0.60184518E-06 \\
(1 2 0) &   0.16729966E-01 &  -0.86926828E-02 &  -0.47449408E-02 &  -0.25011753E-03 &   0.52349008E-04 &   0.39216750E-05 \\
(1 2 1) &  -0.68420124E-02 &   0.15559548E-02 &  -0.95408753E-03 &  -0.59427782E-03 &  -0.82159092E-04 &  -0.33850261E-05 \\
(1 2 2) &   0.80345403E-03 &  -0.54456722E-04 &   0.18658915E-03 &   0.84258499E-04 &   0.10778557E-04 &   0.42789250E-06 \\
(1 3 0) &   0.22208796E-02 &  -0.58431586E-03 &  -0.48924660E-03 &  -0.27928743E-04 &   0.51958515E-05 &   0.39174790E-06 \\
(1 3 1) &  -0.38128951E-04 &   0.13477913E-03 &  -0.25264265E-03 &  -0.10906523E-03 &  -0.13641539E-04 &  -0.53623767E-06 \\
(1 3 2) &   0.87033985E-05 &  -0.14763804E-04 &   0.25957223E-04 &   0.11060270E-04 &   0.13830492E-05 &   0.54549206E-07 \\
(2 0 0) &  -0.35795231E+00 &   0.65313508E-01 &  -0.50976726E-02 &  -0.52759476E-02 &  -0.70390662E-03 &  -0.28026723E-04 \\
(2 0 1) &  -0.80001226E-01 &  -0.34222505E-01 &   0.15934024E-02 &   0.15341728E-02 &   0.17475526E-03 &   0.60919963E-05 \\
(2 0 2) &   0.87639890E-02 &   0.34113860E-02 &  -0.18169113E-03 &  -0.16835979E-03 &  -0.18690148E-04 &  -0.62524682E-06 \\
(2 1 0) &  -0.26504203E-01 &  -0.19821713E-02 &   0.26199589E-02 &   0.52733957E-03 &   0.46965191E-04 &   0.16425008E-05 \\
(2 1 1) &   0.31124751E-03 &   0.71374069E-02 &   0.19284097E-02 &   0.75522804E-04 &  -0.14569881E-04 &  -0.96859723E-06 \\
(2 1 2) &   0.55387018E-03 &  -0.72916289E-03 &  -0.22397964E-03 &  -0.51781606E-06 &   0.30976122E-05 &   0.17280133E-06 \\
(2 2 0) &   0.25359898E-02 &  -0.29959443E-02 &  -0.89093469E-03 &   0.56975030E-04 &   0.25922842E-04 &   0.14212540E-05 \\
(2 2 1) &  -0.77368564E-03 &   0.92068559E-03 &  -0.29143432E-03 &  -0.18699531E-03 &  -0.25019943E-04 &  -0.10043456E-05 \\
(2 2 2) &   0.16753641E-03 &  -0.63376050E-04 &   0.32826799E-04 &   0.18838799E-04 &   0.24833739E-05 &   0.98856519E-07 \\
(2 3 0) &   0.37522544E-03 &  -0.17359581E-03 &  -0.84952992E-04 &   0.16008488E-05 &   0.19086792E-05 &   0.11081771E-06 \\
(2 3 1) &   0.77690104E-04 &   0.52930691E-04 &  -0.71571141E-04 &  -0.28796227E-04 &  -0.34496071E-05 &  -0.13183005E-06 \\
(2 3 2) &  -0.33219791E-05 &  -0.36269908E-05 &   0.64832962E-05 &   0.24768957E-05 &   0.29318259E-06 &   0.11173399E-07 \\
(3 0 0) &  -0.26370968E-01 &   0.45853683E-02 &  -0.42857969E-03 &  -0.38478482E-03 &  -0.50492454E-04 &  -0.19949740E-05 \\
(3 0 1) &  -0.59333101E-02 &  -0.24985017E-02 &   0.83613754E-04 &   0.93963534E-04 &   0.10360895E-04 &   0.34793936E-06 \\
(3 0 2) &   0.82864537E-03 &   0.29126396E-03 &   0.91218775E-05 &  -0.58268328E-05 &  -0.64542109E-06 &  -0.18840624E-07 \\
(3 1 0) &  -0.21245015E-02 &  -0.24321040E-03 &   0.27708187E-03 &   0.76882914E-04 &   0.81206123E-05 &   0.30009780E-06 \\
(3 1 1) &   0.21504225E-03 &   0.58749970E-03 &   0.68725126E-04 &  -0.23819124E-04 &  -0.45232048E-05 &  -0.20090275E-06 \\
(3 1 2) &   0.13663304E-04 &  -0.67148914E-04 &  -0.80040661E-05 &   0.38261029E-05 &   0.69827212E-06 &   0.30722829E-07 \\
(3 2 0) &   0.86363838E-04 &  -0.25412269E-03 &  -0.47306183E-04 &   0.10707095E-04 &   0.25867542E-05 &   0.12762594E-06 \\
(3 2 1) &   0.69686411E-05 &   0.95111050E-04 &  -0.24784177E-04 &  -0.15707547E-04 &  -0.20542771E-05 &  -0.81082411E-07 \\
(3 2 2) &   0.86898718E-05 &  -0.74726285E-05 &   0.15429839E-05 &   0.12006276E-05 &   0.16251525E-06 &   0.64782353E-08 \\
(3 3 0) &   0.18940167E-04 &  -0.14961780E-04 &  -0.47045372E-05 &   0.55656921E-06 &   0.17796779E-06 &   0.91749454E-08 \\
(3 3 1) &   0.10317043E-04 &   0.54645836E-05 &  -0.52425736E-05 &  -0.20926819E-05 &  -0.24595118E-06 &  -0.92493584E-08 \\
(3 3 2) &  -0.50524289E-06 &  -0.31899297E-06 &   0.39480448E-06 &   0.14383851E-06 &   0.16356627E-07 &   0.60444987E-09 \\
 \hline
 \end{tabular}
 }
 \end{table}

\clearpage


\clearpage

\clearpage

\clearpage

\clearpage

\clearpage

\clearpage
\end{document}